\documentclass{iopconfser}
\usepackage{graphicx}
\usepackage{amsmath}
\usepackage{float}
\usepackage{subfigure}
\usepackage{xcolor}
\usepackage{verbatim}

\begin{document}

\title{Low Latency, High Bandwidth Streaming of Experimental Data with EJFAT}

\author{%
Ilya Baldin$^{1}$, Michael Goodrich$^{1}$, Vardan Gyurjyan$^{1}$, Graham Heyes$^{1}$, Derek Howard$^{2}$, Yatish Kumar$^{2}$, David Lawrence$^{1}$, \\
Brad Sawatzky$^{1}$, Stacey Sheldon$^{2}$ and Carl Timmer$^{1}$}

\affil{$^1$Thomas Jefferson National Accelerator Facility, 12000 Jefferson Avenue, Newport News, VA}
\affil{$^2$Energy Sciences Network, LBNL, 1 Cyclotron Road Mail Stop 59R3101 Berkeley, CA  94720}

\email{goodrich@jlab.org}

\begin{abstract}
Thomas Jefferson National Accelerator Facility (JLab) has partnered with Energy Sciences Network (ESnet) to define and implement an edge to compute cluster computational load balancing acceleration architecture. The ESnet-JLab FPGA Accelerated Transport (EJFAT) architecture focuses on FPGA acceleration to address compression, fragmentation, UDP packet destination redirection (Network Address Translation (NAT)) and decompression and reassembly. 

EJFAT seamlessly integrates edge and cluster computing to support direct processing of streamed experimental data. This will directly benefit the JLab science program as well as data centers of the future that require high throughput and low latency for both time-critical data acquisition systems and data center workflows. 

The EJFAT project will be presented along with how it is synergistic with other DOE activities such as an Integrated Research Infrastructure (IRI), and recent results using data sources at JLab, an EJFAT LB at ESnet, and computational cluster resources at Lawrence Berkeley National Laboratory (LBNL).

\end{abstract}

\section{Introduction}

Advances in computing technologies, including FPGA capacities, GPUs and
heterogeneous architectures have opened the door to a major shift in
experimental design and detector readout philosophy\footnote{This material is based upon work supported by the U.S. Department of Energy, Office of Science, Office of Nuclear Physics under contract DE-AC05-06OR23177.}.

Historically, technological constraints meant that the prior generation of DAQ
front-end modules/digitizers were intrinsically `science/physics event based' readout systems.
Digitization of a detector response was predicated on an electronic `trigger'
signal arriving at each module to initiate digitization and logging of timing,
accumulated energy, or similar detector response over some pre-programmed time
period characteristic of the `science event' of interest.  During this measurement
period, the DAQ system was busy and generally blind until the digitization was
complete, the trigger reset, and the process could repeat.

The trigger, flagging a science event of interest in the detector, was historically
composed from a very limited subset of detector channels.  In order to operate
effectively, it needed to be very low-latency, very well understood, and and
extremely reliable.  In the NP/HEP domain this historically involved
implementing a complex logic circuit using thousands of physical cables,
discrete logic modules, and other supporting electronics.

As technology progressed, the physical wiring and racks of discrete electronics
that embodied the trigger logic could be replaced by ASICs and/or FPGAs.  This
miniaturization allowed for much more complex detector combinatorics to be
involved in the trigger decision, significantly improving signal:noise in the
science events selected for readout, but the trigger decision was still bound to custom
hardware.

The next-generation DAQ framework intends to decouple `science event identification'
from the digitization and readout process entirely.  Instead of predicating
detector readout upon identifying a `science event of interest' first, the front end
digitizers continuously stream measurement data from the detector to a 

\textcolor{black}{consumption}
layer based on high-performance, but otherwise commodity hardware for processing.
This \textcolor{black}{consumption}-layer can have access to the full detector response all the time
and is vastly more capable than historical hardware or firmware based solutions.
This ``Streaming Read Out'' (SRO) model has many advantages:
\begin{itemize}
    \item All detector channels may be part of ``the trigger'', allowing for
        much more sophisticated and efficient selection of  data  stored
        for future processing.
    \item Multiple ``triggers" can be implemented in parallel and on the fly as
        the readout layer, processing layers, and storage decisions are now
        cleanly decoupled.
  \item Detector signal processing can now be implemented in the middle layer
       via modern high-level programming languages and software
        development tool kits.
        This allows for detector monitoring, calibrations, and slow-control
        detector tuning to be done in near-realtime.
  \item Custom trigger hardware and firmware requiring specialized knowledge and tools
      is minimized.
  \item Enables sophisticated tagging/filtering algorithms limited only by the
      (scalable) available computing power.
  \item The flexible, software-defined, middle layer can take best advantage of
      emerging technologies such as AI/ML tools and heterogeneous computing.
  \item All of this leads to a much cleaner/calibrated data stream from the
      detectors, improved signal:noise in the stored data, and ultimately an accelerated
        path to physics output.
\end{itemize}

However, all of this high-level discussion on accessing and processing the raw
data stream is predicated on a \emph{robust, low latency, high bandwidth
environment} that can transparently scale to support the extraordinary data
volumes delivered by modern front-end digitizers.  Moreover, the underlying architecture
should be able to dynamically scale resources to meet demand on the fly.  If beam current
rises and the \textcolor{black}{consumption}-layer requires more compute power (or a different mix of
resources), then the environment must be capable of bringing those resources online and
\emph{transparently} handling the needed data duplication and/or rerouting without impacting
data collection on the front end.

This increasingly complex data transport `fabric' is itself subject to to
challenges of load and isolated component failures.  In order for it to serve
the needs of any complex future experiment, this absolutely critical transport
layer must be adaptive, capable of `self-healing' and be able to transparently
redirect data flows in response to demand and/or isolated failures.

\section{Accelerated Load Balanced Transport Fabric}
EJFAT is designed to stream real time DAQ samples across the wide area network from a source site to one or more receiving sites.  In order to perform this task, it was designed with the following assumptions:
\begin{itemize}
    \item Unidirectional streaming at terabits per second.
    \item Secure tunneling of UDP packets between sites, with an operationally friendly model that allows DAQs to push data, and Compute Nodes (CNs)to register for service (pull).
    \item Coherent aggregation of \emph{events}\footnote{Data aggregation event (often a simple time-slice) sent to a single receiving host to be processed such that e.g., zero or more \emph{physics} events may be discovered therein.} to  CNs whilst being robust to packet reordering, where \emph{event} is context defined, but always constitutes an aggregation event to EJFAT.
    \item Dynamic control of event rates, based on feedback from CNs on event consumption rates.
    \item Seamless addition and deletion of CNs without any packet loss.
    \item Use case dependent acceleration opportunities for source to EJFAT data preparation, UDP fragmentation, and CN reassembly.
\end{itemize}

For more technical details on the EJFAT system, refer to previous paper [1].

\begin{figure}[H]
    \centering
    \includegraphics[width=0.95\linewidth]{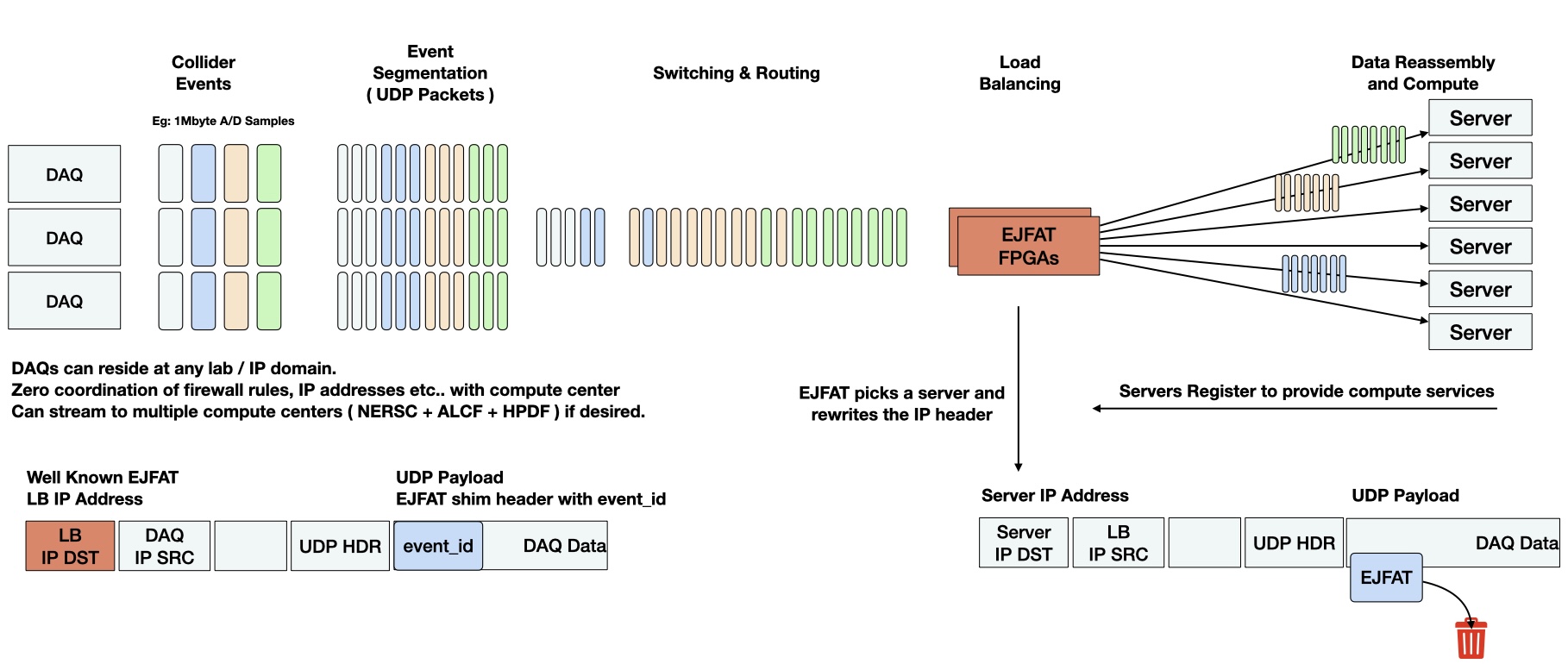}
    \caption{EJFAT End to End Packet Flow and Processing}
    \label{fig:EJFAT-overview}
\end{figure}

\subsection{Source Data Preparation}

Figure (\ref{fig:EJFAT-overview}) is a packet level view of the EJFAT acceleration fabric.  'Event' packets are emitted by different portions of a large scientific instrument (e.g., sub-detectors or channels of a detector) with a common aggregation tag in included meta-data  and additional meta-data for individual and parallelizable channel reassembly by a  CN. After proper meta-data tagging, packets are forwarded to the well known LB IPV4/6 address and individually re-directed to some CN using the aggregation tag at line rate (100Gbps or more) with microsecond level latency by the EJFAT FPGA based Data Plane (DP) using a NAT like network function  using a dynamically weighted round robin algorithm configured by the CP and updated to the FPGA DP at a 1 Hz rate.

The EJFAT system understands 'event' as an \emph{data aggregation event}  and leaves any consideration of science/physics event entirely to back end processing.  For work to  be load balanced across a cluster of CNs by the  EJFAT architecture, the data stream must be sent as a sequence of data aggregation events. Each data aggregation event is  processed as a meaningful whole by a single receiver CN.  Thus the EJFAT DP fulfills the event aggregation function present in previous DAQ designs.

\subsection{LB dataplane forwarding}

For operational simplicity, the load balancer takes ownership of dynamically mapping the   DAQ packets destination IP address to specific addresses not known a priori to the data source and dynamically determined by the subscription process.  This has the property that DAQs simply need to send all packets to a single well known IP address for the load balancer.  The secure network connection between the DAQ and the load balancer can be integrated once, regardless of the final selection of computational facility.   As well, computational facilities do not get any work pushed into them, instead they register and withdraw CNs for service with the load balancer.  This reception side network integration is also done once between the load balancing host and the compute center, rather than once per experiment or DAQ facility.

Once the packets are flowing, the CNs provide 1 second updates on queue fill levels back to the FPGA based load balancer.  The load balancer operates a feedback control loop to increase or decrease the event rate for each individual CN.  Terminating the control loop at the load balancer removes the burden of implementation at the DAQ sources, which are then free to run at maximum speed with no backpressure.

A CN is deleted from the pool by requesting a free operation (or alternatively added via subscription) from the EJFAT \emph{control plane} (CP), where it is immediately removed from the pool for new data aggregation events.  However it continues to remain in the pool for data aggregation events that are in progress.  After a reasonable delay, it is removed entirely, allowing data aggregation event traffic to arrive without packet loss at the boundary.

A further level of complexity occurs when data aggregation events are comprised from multiple channels (i.e. different parts of the same scientific instrument) and must be reassembled and processed on the same host. \textcolor{black}{Thus in addition to tagging data from related sources with the same data aggregation tag, the meta data contains a channel (e.g., subdetctor) tag which directs the LB to send event packets to distinct UDP ports at the receiving CN. A significant benefit of this design is the parallel processing that results}.

On a practical note, currently all data aggregation event reassembly is done in software. While it's possible to work hard to recover UDP packets which are significantly out of sequence, such code creates enough of a burden to cause many more packet drops in high speed applications than are ever recovered. We found performance to be best when ignoring such packets.  Alternatively, further acceleration opportunities may be pursued, e.g. FPGA based decompression and (possibly out-of-sequence) reassembly.

For time-critical workflows (e.g. DAQ applications), it is possible that the data aggregation event fragmenting/reassembly functions will be absorbed into modified existing hardware/firmware or be placed into additional FPGAs. It should be noted that LB operations are only dependent on proper data aggregation event tagging of packets and therefore EJFAT may be used for both free streamed and triggered data aggregation event sources.

\subsection{LB Control Plane}

The EJFAT CP is built on top of the ESnet SmartNIC [5] platform, and is responsible for dynamically re-configuring DPs. It supports reserving virtual load balancer instances on the FPGA(s), each of which individually support dynamically adding and removing senders and receivers (CNs). As the receivers receive data aggregation events and place them on a queue, the queue level is measured in a Proportional, Integral, Derivative (PID) controller loop. The control variables from each receiver are sent to the EJFAT CP to adjust the proportion of data aggregation events that it will receive each second. This allows the CP to adapt the data plane to heterogeneous performing receivers with variable service rates, load balance data aggregation event rates across registered nodes, dynamically weave spontaneous new or returning registrants into the load balance, and evict retiring nodes.

In order to make these changes without redirecting data aggregation events in-flight, the CP requires knowledge of the current data aggregation tag. Senders provide this knowledge to the CP by sending data aggregation tag sync messages, which the CP uses to build a linear model. This model predicts future data aggregation tags, so that the CP can make changes that won't affect in-flight traffic. 

CP subscribers (CNs) provide network coordinates and port ranges for the CP to configure look up tables (LUTs) in the DP for the DP's core NAT functionality.

A single CP can support up to 8 virtual LB instances where each virtual LB instance constitutes an independent experiment and each virtual LB instance DP  can be serviced by 1 or more FPGA cards. This facilitates experiment data rate scaling in 200Gbps increments (the max for an single FPGA) to 1 Tbps and beyond.

The CP also provides a dashboard, Prometheus metrics, and gRPC procedures to query the immediate state of an LB instance. The state of the CP is persisted in a SQLite database, allowing for crash recovery and upgrades.

\section{EJFAT Demonstration}
In a demonstration of the EJFAT system's capabilities, a dataset from a CLAS12 experiment was replayed as if it were streaming directly from the data acquisition system during live experiments. This transition from traditional file/batch processing to a streaming processing paradigm was undertaken to rigorously test the EJFAT system's proficiency in real-time data stream processing applications. The operational CLAS12 reconstruction application, initially designed and executed within the CLARA framework [2], was modified and re-implemented within the ERSAP framework. This modification was crucial in developing a streaming, real-time physics event reconstruction application that leveraged existing CLAS12 subdetector reconstruction algorithms.

\begin{figure}[H]
    \centering
    \includegraphics[width=0.8\linewidth]{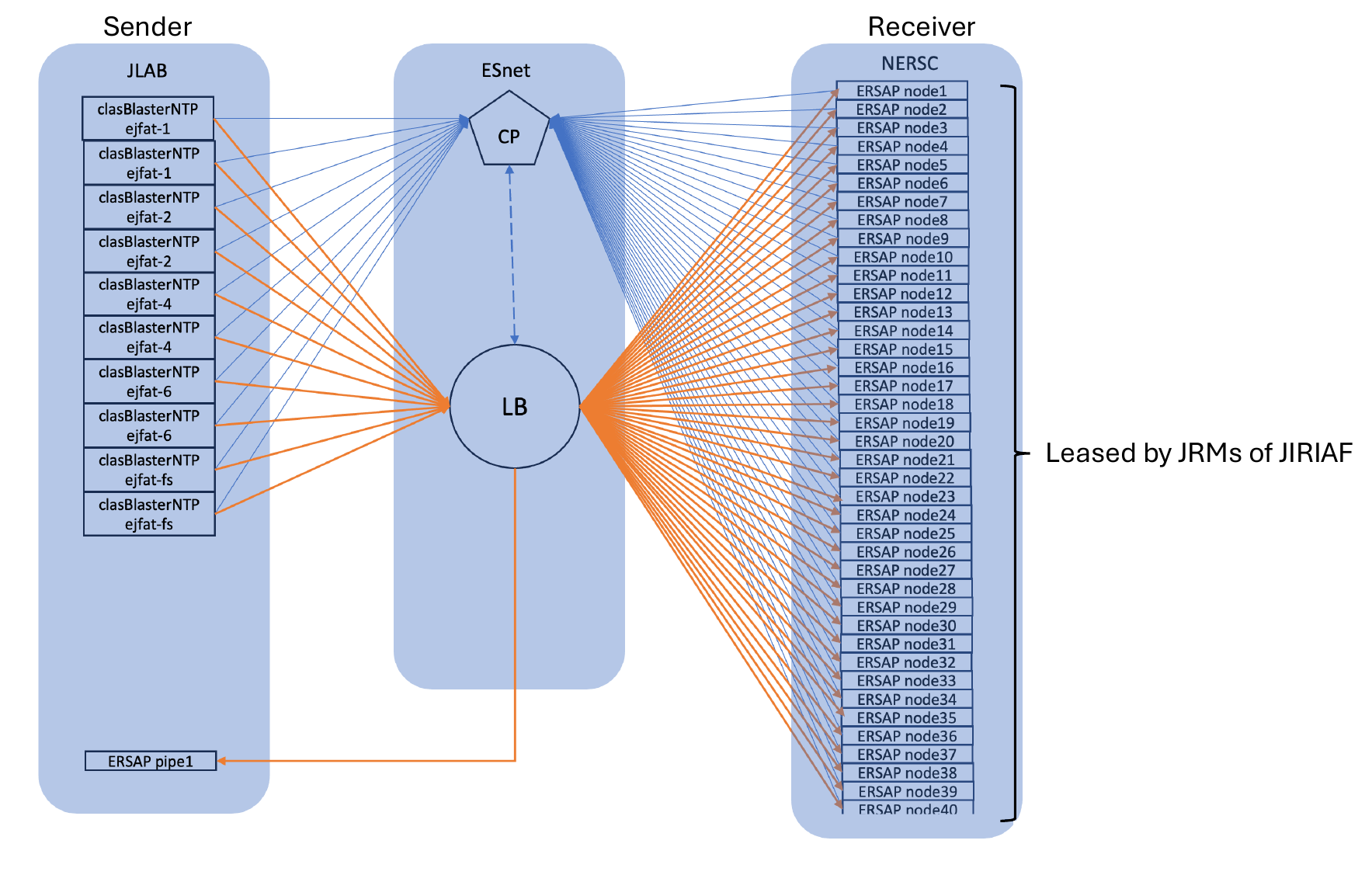}
    \caption{CLAS12 Remote Data Stream Processing Data-Flow Diagram. 
The EJFAT packetizer segments data aggregation events into UDP packets, facilitating their transport. The EJFAT reassembly engine reconstructs these UDP packets into a cohesive data aggregation event for backend processing. The nodes in the ERSAP data pipeline represent CLAS12 reconstruction actors that process data corresponding to specific CLAS12 detector components. Blue lines are control and orange lines data paths.  }
    \label{fig:C12RDS-DFD}
\end{figure}

\begin{figure}[H]
    \centering
        \includegraphics[width=0.65\textwidth]{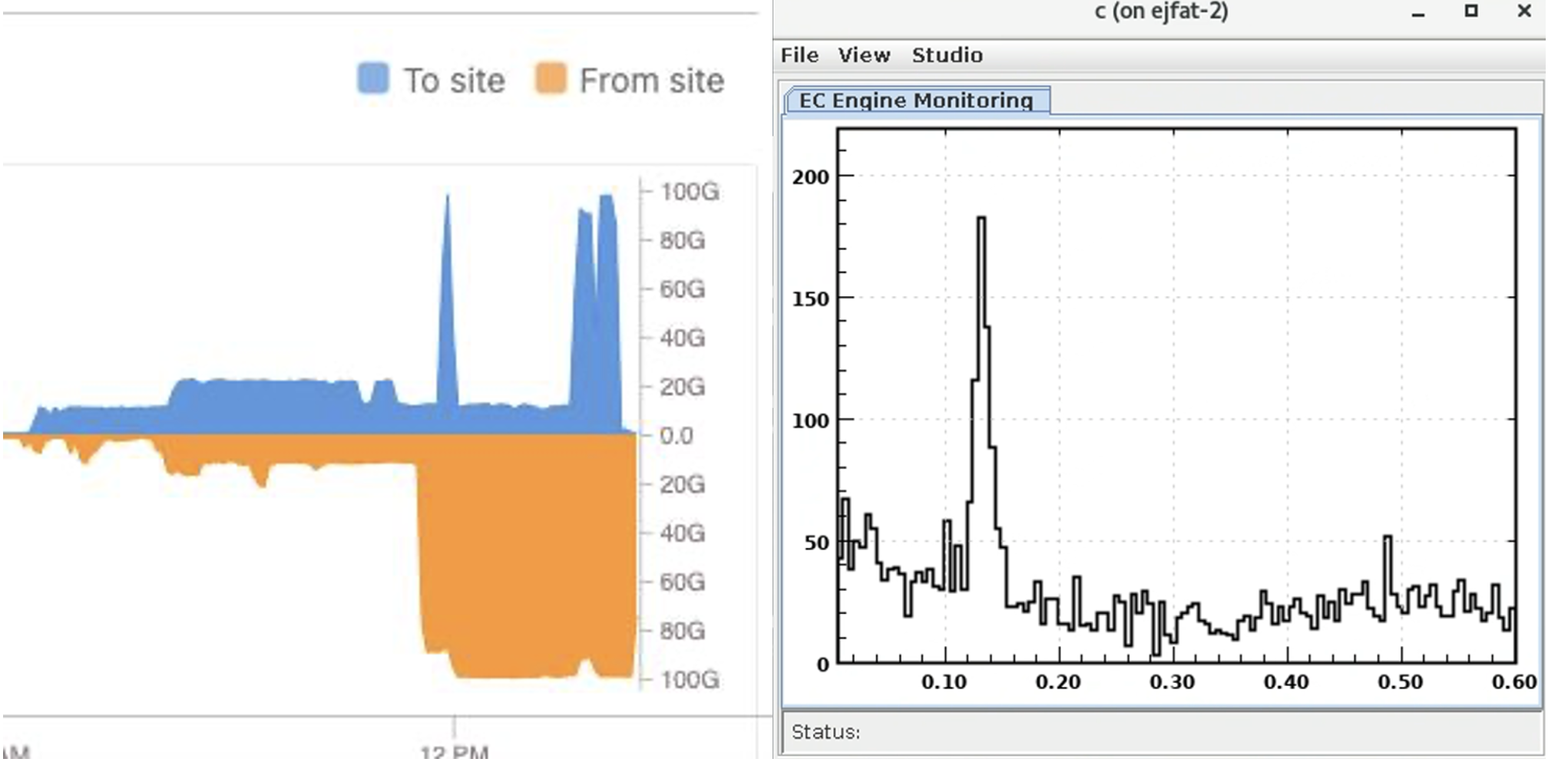}
    \caption{These plots illustrate data throughput into and out of ESnet during the concept validation experiment, and a plot from the data processing validation highlighting the pi0 missing mass reconstruction.  }
    \label{fig:twofigures}
\end{figure}

Each component's reconstruction algorithms within the CLAS12 detector were encapsulated as ERSAP data-processing reactive actors [3]. These actors were interconnected to form comprehensive applications capable of accurately identifying particles and determining their momenta, trajectories, and energies before and after interactions. The precision and accuracy of this reconstructed data were subsequently verified by a physics validation actor, ensuring the reliability of the reconstruction application.
In a proof-of-concept data stream processing experiment, raw experimental data from CLAS12 was successfully streamed to the National Energy Research Scientific Computing Center (NERSC) through the EJFAT load balancer at the Energy Sciences Network (ESnet) at impressive rates exceeding 100 Gbps. This data underwent massive parallel processing on 40 Perlmutter nodes, utilizing over 10,000 cores, and was then streamed back to Jefferson Lab in real time for validation, persistence, and final physics analysis. Figure (\ref{fig:twofigures}) illustrates a 100 Gbps data flow through the EJFAT LB, as monitored by ESnet.

This innovative data processing pipeline is significant as it eliminates intermediate data storage along the data-stream path, with data originating and being stored exclusively at Jefferson Lab. This approach effectively reduces file I/O latencies, enhancing overall processing efficiency.
For this experiment, the JIRIAF system [4] played a critical role in managing computing resources, deploying workflows, and orchestrating the overall process. 

\section{DOE Synergies}
The Integrated Research Infrastructure (IRI) initiated by DOE ASCR seeks ‘to empower researchers to meld DOE’s world-class research tools, infrastructure, and user facilities seamlessly and securely in novel ways to radically accelerate discovery and innovation’ [6]. IRI enables three types of science patterns: time sensitive - requiring urgency and real-time access to compute resources and instruments, data integration, which combines and computes on data from multiple sources and long-term campaigns, which requires continuous sustained access to compute resources over a long period of time. Of these, EJFAT fits with and enables the first pattern – time sensitive. The types of science that benefit from using EJFAT are precisely the ones that work in real-time:  EJFAT architecture is bufferless thus minimizing delays between data being produced and processed, affording significant flexibility in which compute resources are engaged in processing specific event-streams, flexibly adding and removing compute resources from the workflow without loss of data, thus addressing one of the main obstacles to adopting real-time processing – the end-to-end resilience of the system to temporary disruptions. EJFAT is also flexible in its definition of event streams, thus allowing multiple different types of processing pipelines to take advantage of it. EJFAT team is working with the IRI program to help demonstrate the benefits and advantages of this architecture to different DOE science communities.

\section{Summary and Conclusions}

The EJFAT acceleration fabric is well designed to support current and future Tbps and beyond level streaming or triggered workflows.  We anticipate that moving triggers from the data source (e.g., detector) edge to the compute cluster will yield higher quality analysis via full data access and dynamically expandable resources improving the quality of results, and saving analyst effort thus compressing the timeline to publishable results. EJFAT continues to be a key initiative to accelerating the rate at which science progresses.

\section*{References}
\begin{enumerate}
    \item M. Goodrich, C. Timmer, V. Gyurjyan, D. Lawrence, G. Heyes, Y. Kumar, and S. Sheldon, "ESnet/JLab FPGA Accelerated Transport," IEEE Transactions on Nuclear Science, vol. 70, no. 6, Feb. 2023. DOI: 10.1109/TNS.2023.3243871.
\item V. Gyurjyan, et al., "CLARA: A Contemporary Approach to Physics Data Processing," J. Phys.: Conf. Ser., vol. 331, p. 032013, 2011. DOI: 10.1088/1742-6596/331/3/032013.
\item V. Gyurjyan, D. Abbott, 
    N. Brei, M. Goodrich,
    G. Heyes, E. Jastrzembski, D. Lawrence, 
    B. Raydo, and C. Timmer, 
    "ERSAP: Toward Better NP Data-Stream 
    Analytics 
    With Flow-Based Programming," 
    IEEE Transactions on Nuclear Science, 
    vol. 70, no. 6, Jun. 2023. 
\item V. Gyurjyan, C. Larrieu, G. Heyes, and D. Lawrence, "Jiriaf: Jlab integrated research infrastructure across facilities," in EPJ Web of Conferences, vol. 295, p. 04027, 2024.
\item https://github.com/esnet/esnet-smartnic-hw
\item DOE IRI ABA https://www.osti.gov/biblio/1984466
\end{enumerate}

\end{document}